\documentclass[journal]{IEEEtran}

\usepackage{amsmath,amssymb,bm}
\usepackage{cite}
\usepackage{graphicx}
\usepackage{url}

\graphicspath{{img/}}

\begin{document}
\bstctlcite{IEEEcontrol}

\title{Finite-Element Electromagnetic Analysis and Evaluation of Antenna Performance Metrics Using a Quantum Linear System Solver}

\author{Hiroaki Ominato, Takahiro Ohyama, and Shiro Kawabata%
\thanks{Corresponding author: Hiroaki Ominato.}%
\thanks{Hiroaki Ominato is with Panasonic System Networks R\&D Lab. Co., Ltd., Yokohama, Kanagawa 224-8539, Japan (e-mail: ominato.hiroaki@jp.panasonic.com).}%
\thanks{Takahiro Ohyama is with Panasonic System Networks R\&D Lab. Co., Ltd., Sendai 981-3206, Japan (e-mail: ohyama.takahiro-pmcs@jp.panasonic.com).}%
\thanks{Shiro Kawabata is with the Faculty of Computer and Information Sciences, Hosei University, Koganei, Tokyo 184-8584, Japan (e-mail: kawabata@hosei.ac.jp).}}

\maketitle

\begin{abstract}
Electromagnetic analysis based on the finite element method is widely used in antenna design. At higher frequencies, however, the required mesh refinement makes solving large sparse linear systems the main contributor to computational cost. Quantum linear system solvers offer a potential means of accelerating this step, but their output is a superposition state: reconstructing all components of the classical solution vector would forfeit the quantum speedup. Prior studies have proposed methods for obtaining physical quantities such as scattering cross sections and scattering parameters from the solution state. To support antenna design using fault-tolerant quantum computers, we formulate the total radiated power, the radiated power within a prescribed solid angle, and the dielectric and conductive losses as observables for efficient evaluation from a quantum solution state. These observables avoid the measurement cost of reconstructing all components of the output quantum state. We develop an end-to-end hybrid quantum--classical workflow that combines finite-element electromagnetic analysis, matrix preconditioning, a quantum linear system solver, and the evaluation of antenna design metrics. To validate the workflow, we performed numerical experiments evaluating these observables for an axisymmetric coaxially fed monopole antenna. Under idealized conditions, the relative errors in the total radiated power and the dissipated power with respect to the classical finite-element solution were on the order of $10^{-3}$, confirming the consistency of the workflow. The proposed method is expected to provide a basis for using quantum linear system solvers to evaluate antenna performance and, in the future, to design high-frequency wireless devices.
\end{abstract}

\section{Introduction}
\label{sec:introduction}

Antennas for wireless communication devices must meet performance requirements within limited mounting space while accommodating increasing communication capacity, device miniaturization, and multifunctionality. In the high-frequency bands used for 5G and 6G, the electromagnetic characteristics depend not only on the antenna shape but also on the feed structure, substrate, housing, surrounding components, and material losses. Designing an antenna in isolation from its surrounding structures is therefore difficult, making electromagnetic analysis of the antenna within the product increasingly important. Design parameters are adjusted to meet requirements for gain, radiation efficiency, and directivity. A single analysis does not necessarily yield a design that satisfies these requirements simultaneously, so parameter modification, electromagnetic analysis, and performance evaluation are repeated to compare candidates. The overall computational cost of antenna design thus depends on both the analysis cost for a single design candidate and the number of iterations, including frequency sweeps, parameter sweeps, and optimization.

The finite element method (FEM) is widely used for antenna design and verification because it can handle complex materials and boundary geometries. At higher frequencies, shorter wavelengths require finer meshes to represent spatial variations in the electromagnetic field. Including the substrate, housing, and installation environment also enlarges the analysis domain. Both factors increase the number of FEM degrees of freedom, making the solution of the discretized linear system the main contributor to computation time and memory requirements. The coefficient matrix of this linear system is generally sparse, and efficient solution methods that exploit this structure are required. Quantum linear system solvers (QLSSs), beginning with the Harrow--Hassidim--Lloyd (HHL) algorithm \cite{Harrow2009}, offer a way to prepare a quantum state encoding the solution. Subsequent improvements in solution-state precision \cite{ChildsKothariSomma2017} and the matrix-inversion framework of quantum singular value transformation (QSVT) \cite{Gilyen2019} have strengthened the motivation to investigate this approach for electromagnetic analysis.

The usefulness of a QLSS for antenna design depends on what can be extracted from its output. The solver prepares the normalized state $|x\rangle=\bm x/\|\bm x\|$, rather than a classical list of field coefficients. Reconstructing that list incurs a cost that grows with its length, so an efficient application must instead extract selected performance metrics directly. Quantum FEM complexity analyses also assess the cost of obtaining a requested functional at a prescribed accuracy, including discretization and readout \cite{MontanaroPallister2016}. The relevant task is therefore to connect the solution state to the physical quantities needed for design.

Quantum electromagnetic research provides several routes to preparing such field states. Early work used the transmission line matrix method for quantum electromagnetic simulation \cite{sinha2010quantum}; subsequent time-domain approaches developed wave-equation simulation and finite-difference circuits \cite{CostaJordanOstrander2019,SatoEtAl2024}. For Maxwell's equations, quantum time evolution has been formulated through Schr\"odingerisation, including boundaries, interfaces, and time-dependent sources \cite{jin2024maxwell,ma2024sources}, and through electromagnetic potentials \cite{tezuka2026electromagnetic}. Frequency-domain approaches instead target a steady-state field by solving a discretized linear system, as in HHL-based electromagnetic FEM \cite{zhang2024quantum} and QSVT for one-dimensional dissipative electromagnetic problems \cite{novikau2023simulation}. Recent comparisons of HHL and QSVT extend this investigation to Maxwell problems motivated by industrial applications \cite{Turro2026}. This work focuses on the frequency-domain linear-systems formulation, with antenna performance evaluated from the resulting FEM coefficient vector.

Extracting physical quantities from a quantum state already has precedents: energy can be measured in simulations of coupled classical oscillators \cite{Babbush2023} and lossless wave equations \cite{bosch2025sources}, while quantum FEM has yielded S-parameter magnitudes \cite{11103791,11230107} and, more recently, both magnitude and phase \cite{11456245}. For radiation, a particularly close precedent is the radar-cross-section calculation of Clader et al., which uses the squared magnitude of a far-field linear functional \cite{Clader2013}. That squared amplitude is itself a quadratic form of the solution coefficients. Evaluating antenna power requires integrating radiation over the full sphere or a specified angular region and accounting separately for power dissipated in the dielectric and conductor. This requires expressing these angular, volume, and surface integrals as observables of the same FEM solution. Their expectation values, combined with the solution norm, yield the corresponding powers.

\begin{figure*}[t]
  \centering
  \includegraphics[width=.84\textwidth]{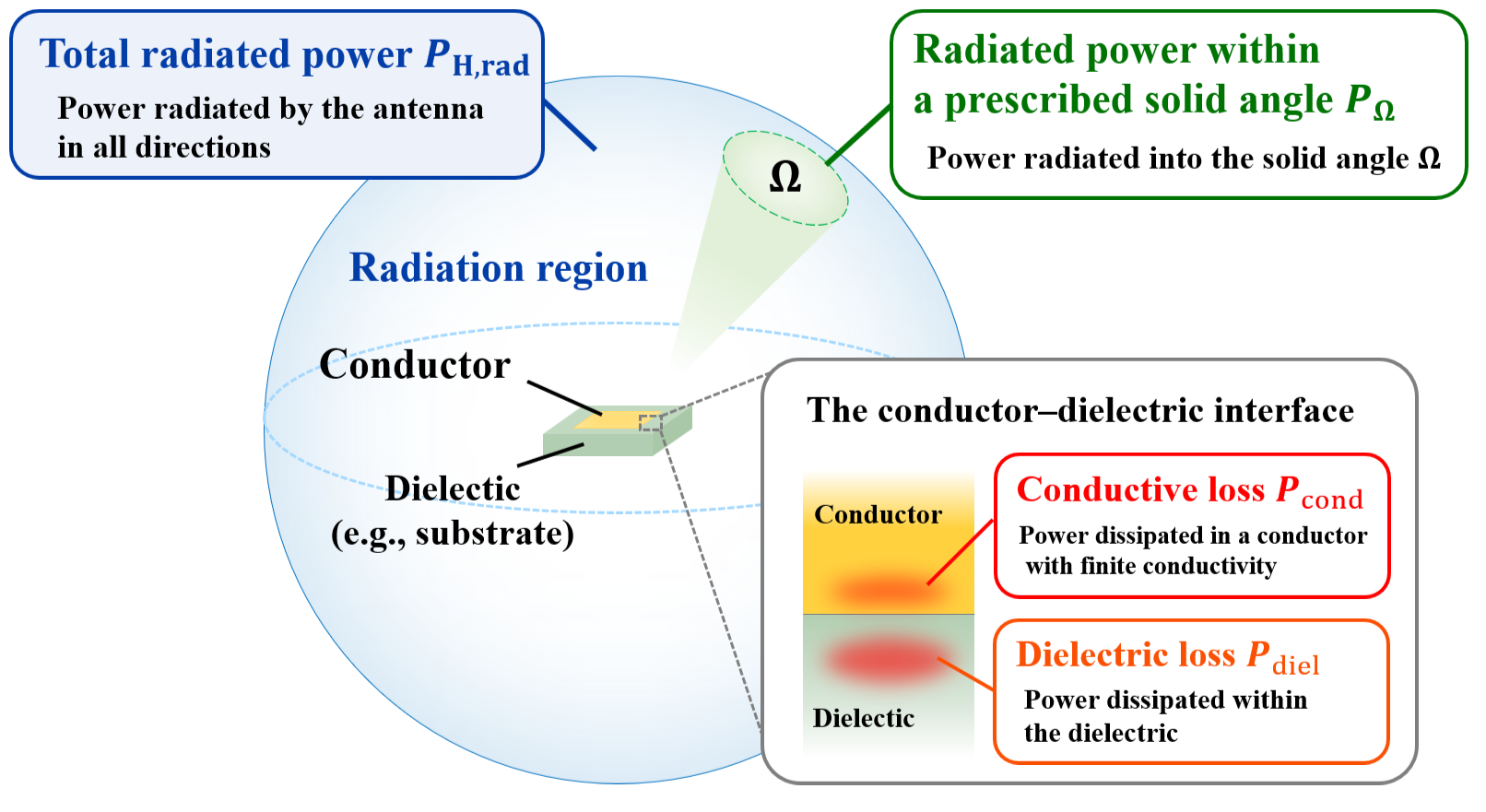}
  \caption{Conceptual diagram of the physical quantities evaluated in this work. Here $P_{H,\mathrm{rad}}$ denotes the radiated power evaluated through the Huygens surface, corresponding to $P_H$ in (\ref{eq:huygens_power}).}
  \label{fig:conceptual_diagram}
\end{figure*}

\begin{figure}[t]
  \centering
  \includegraphics[width=.92\columnwidth]{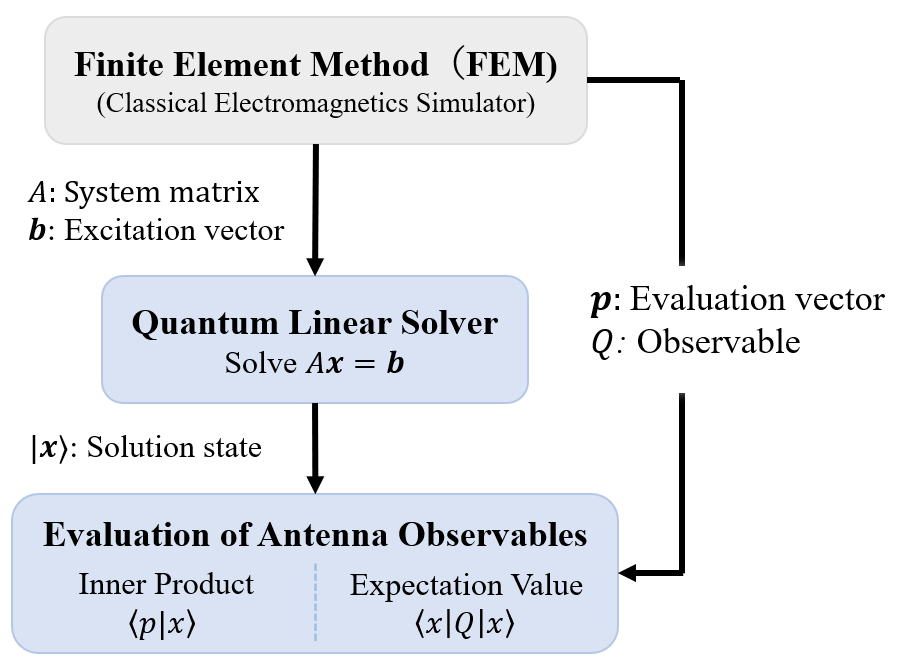}
  \caption{Hybrid quantum--classical workflow for evaluating antenna observables from the finite-element equation.}
  \label{fig:workflow}
\end{figure}

In this work, we construct explicit matrix representations of the total radiated power, the radiated power within a prescribed solid angle, and the dielectric and conductive losses shown in Fig.~\ref{fig:conceptual_diagram}. We derive these observables from integrals of the frequency-domain FEM basis fields and evaluate them through the workflow in Fig.~\ref{fig:workflow}. Comparisons with classical solutions of the same discrete equation verify the consistency of the formulation and workflow.

Section~\ref{sec:fem} presents the finite-element equation and the matrix properties. Section~\ref{sec:observables} formulates the proposed antenna-oriented observables, and Section~\ref{sec:workflow} describes their evaluation using a preconditioned QLSS. Section~\ref{sec:numerical_results} presents the numerical experiments, Section~\ref{sec:discussion} discusses limitations and future work, and Section~\ref{sec:conclusion} summarizes the conclusions.

\section{Finite-Element Electromagnetic Analysis}
\label{sec:fem}
\subsection{Frequency-Domain Formulation}
\label{sec:fem_formulation}

We use the edge-element FEM framework for the frequency-domain Maxwell equations \cite{MonkZhang2019}. To evaluate radiated power and material losses from the same FEM solution, we first specify the treatment of the feed, the open boundary, and losses. In the basic model, the conductors are treated as perfect electric conductors (PEC). In models that include losses, dissipation is represented by the complex permittivity of the dielectric and the surface impedance of the conductor.

We use the time-harmonic convention $e^{-j\omega t}$ and denote the electric field by $\bm{E}$ and the magnetic field by $\bm{H}$. Here $j^2=-1$, $\omega$ is the angular frequency, $\bm\mu$ is the permeability, and $\bm{J}_{\rm imp}$ is the impressed current density. The complex permittivity of a passive dielectric is written as $\widetilde{\bm\varepsilon}=\bm\varepsilon+j\bm\varepsilon_{\rm loss}$ ($\bm\varepsilon_{\rm loss}$ denotes a real coefficient or a real diagonal matrix representing the loss component of the complex permittivity). The electric field is given by
\begin{equation}
 \nabla\times(\bm\mu^{-1}\nabla\times\bm E)-\omega^2\widetilde{\bm\varepsilon}\bm E
 =j\omega\bm J_{\rm imp}. \label{eq:maxwell}
\end{equation}
The PEC condition $\bm n\times\bm E=0$ is imposed on conductor surfaces, where $\bm n$ is the outward unit normal of the electromagnetic analysis domain. When finite conductivity is considered, the PEC condition is replaced by the surface impedance boundary condition (SIBC)
\begin{equation}
 \bm E_t=Z_s(\bm H_t\times\bm n),\qquad
 Z_s=R_s-jX_s,\quad R_s\geq0
 \label{eq:sibc}
\end{equation}
on the conductor surface $\Gamma_s$. The subscript $t$ denotes the tangential component, and $R_s$ and $X_s$ are the surface resistance and surface reactance, respectively. With the unit normal $\bm n_s=-\bm n$ directed outward from the conductor, the same condition reads $\bm E_t=Z_s(\bm n_s\times\bm H_t)$. In accordance with the coaxially fed antenna used in the numerical experiments, the local absorbing condition for the fundamental transverse electromagnetic (TEM) mode at the input cross section $\Gamma_p$ is
\begin{equation}
 \bm n\times\bm H_t+\eta_p^{-1}\bm E_t
 =2a_1\eta_p^{-1}\bm e_1
 \quad\text{on }\Gamma_p,
 \label{eq:port}
\end{equation}
where $\eta_p$ is the wave impedance of the TEM mode, $a_1$ is the known incident amplitude, and $\bm e_1$ is its tangential electric field distribution. The open region is truncated by a perfectly matched layer (PML).

Let $\Omega$ denote the computational domain including the PML; in the following volume integrals, the material tensors after coordinate stretching are used for $\bm\mu,\widetilde{\bm\varepsilon}$ inside the PML. Using a test function $\bm v$ whose tangential component vanishes on the PEC surfaces, integration of (\ref{eq:maxwell}) by parts gives the boundary term $j\omega\int_{\partial\Omega}\bm v_t^*\cdot(\bm n\times\bm H_t)\,dS$. Substituting (\ref{eq:sibc}) and (\ref{eq:port}) into this term yields the weak form
\begin{equation}
 \begin{aligned}
 &\int_\Omega(\nabla\times\bm v)^*\cdot\bm\mu^{-1}(\nabla\times\bm E_h)\,dV
 -\omega^2\int_\Omega\bm v^*\cdot\widetilde{\bm\varepsilon}\bm E_h\,dV\\
 &\quad-j\omega\int_{\Gamma_p}\eta_p^{-1}\bm v_t^*\cdot\bm E_{h,t}\,dS
 -j\omega\int_{\Gamma_s}Z_s^{-1}\bm v_t^*\cdot\bm E_{h,t}\,dS\\
 &=j\omega\int_\Omega\bm v^*\cdot\bm J_{\rm imp}\,dV
 -2j\omega a_1\int_{\Gamma_p}\eta_p^{-1}\bm v_t^*\cdot\bm e_1\,dS .
 \end{aligned}
 \label{eq:weak_form}
\end{equation}
Here $*$ denotes the complex conjugate, and the boundary term on the outer truncation boundary is set to zero. The negative signs of the SIBC and port terms follow from the outward-normal convention, and the factor $2$ in the excitation term follows from (\ref{eq:port}). We expand $\bm E_h=\sum_jx_j\bm N_j$ in $H(\operatorname{curl})$-conforming Nedelec edge basis functions $\bm N_j$ \cite{Nedelec1980} and choose $\bm v=\bm N_i$. The stiffness matrix, mass matrix, port matrix, SIBC boundary matrix, and source vectors are defined as
\begin{equation}
 \begin{aligned}
 K_{ij}&=\int_\Omega(\nabla\times\bm N_i)^*\cdot\bm\mu^{-1}(\nabla\times\bm N_j)\,dV,\\
 M_{ij}&=\int_\Omega\bm N_i^*\cdot\widetilde{\bm\varepsilon}\bm N_j\,dV,\\
 P_{ij}&=\int_{\Gamma_p}\eta_p^{-1}\bm N_{i,t}^*\cdot\bm N_{j,t}\,dS,\\
 [B_{\rm SIBC}]_{ij}&=-j\omega\int_{\Gamma_s}Z_s^{-1}\bm N_{i,t}^*\cdot\bm N_{j,t}\,dS,\\
 f_i&=j\omega\int_\Omega\bm N_i^*\cdot\bm J_{\rm imp}\,dV,\\
 g_i&=\int_{\Gamma_p}\eta_p^{-1}\bm N_{i,t}^*\cdot\bm e_1\,dS
 \end{aligned}
 \label{eq:fem_entries}
\end{equation}
where $\bm N_{j,t}=\bm N_j-(\bm n\cdot\bm N_j)\bm n$. The discrete equation is
\begin{equation}
 \begin{aligned}
 A\bm x&=\bm b,\quad A=K-\omega^2M-j\omega P+B_{\rm SIBC},\\
 \bm b&=\bm f-2j\omega a_1\bm g .
 \end{aligned}\label{eq:linear_system}
\end{equation}
When all conductors are treated as PEC, $B_{\rm SIBC}$ need not be added. Each edge basis function is nonzero only on the adjacent elements that share that edge. Therefore, a matrix entry is nonzero only when the supports of two basis functions overlap on the same element or boundary face, and $K,M,P,B_{\rm SIBC}$, and $A$ are sparse matrices. The sparsity of the right-hand side depends on the basis and excitation settings.

For reciprocal materials and a PML, with real-valued basis functions, the matrix satisfies
\begin{equation}
 A^{\mathsf T}=A,\qquad A^{\dagger}\ne A. \label{eq:complex_symmetric}
\end{equation}
That is, $A$ is complex symmetric and in general neither Hermitian nor positive definite, so solvers that assume these properties cannot be applied directly. Here $\dagger$ denotes the complex conjugate transpose and $\mathsf{T}$ the transpose.

\subsection{Axisymmetric Model}

\begin{figure*}[t]
  \centering
  \includegraphics[width=.86\textwidth]{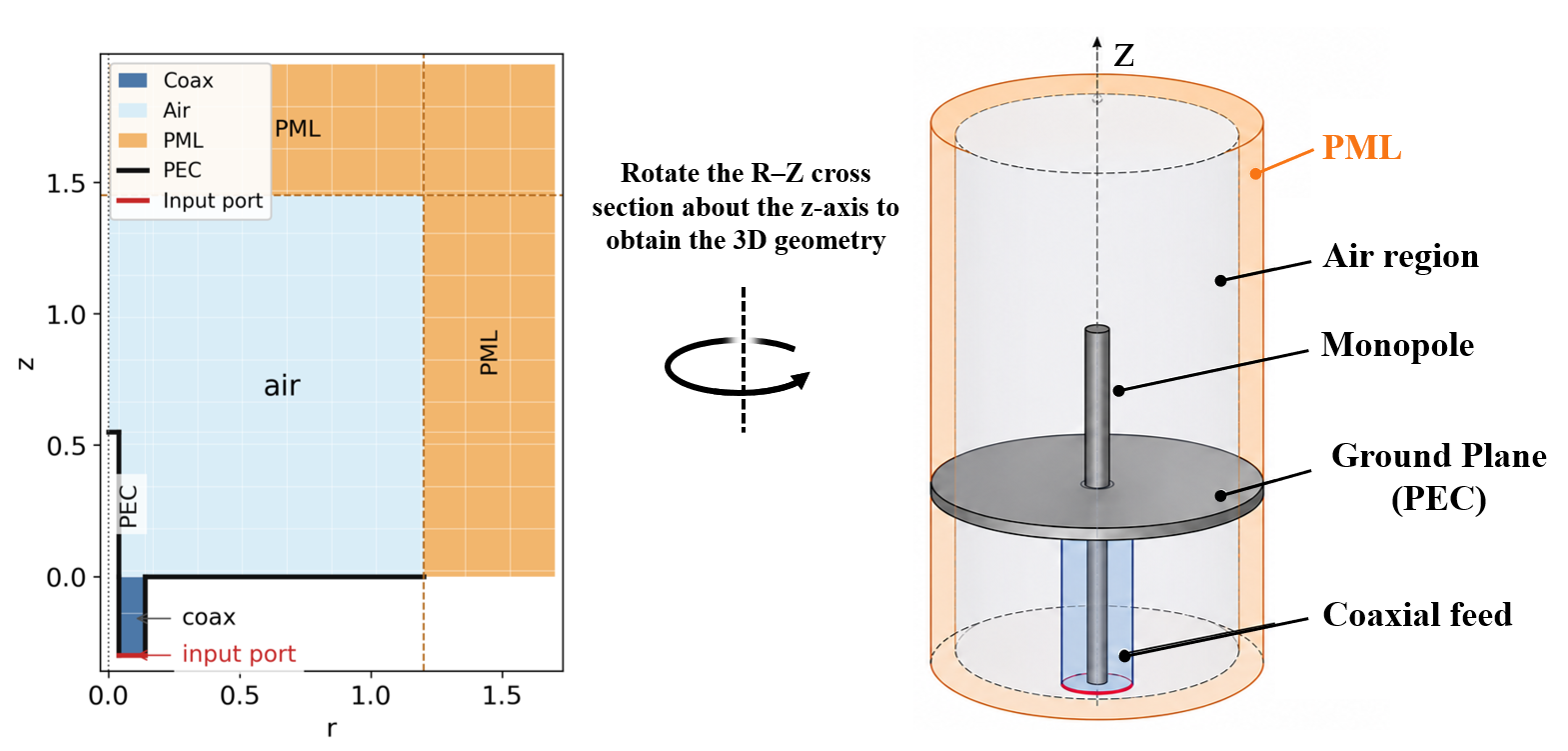}
  \caption{Computational model of the axisymmetric coaxially fed monopole antenna. Left: $R$--$Z$ cross-section mesh; right: the corresponding rotated geometry.}
  \label{fig:model}
\end{figure*}

We use a coaxially fed monopole antenna to verify the observable formulation. The geometry, materials, and excitation are rotationally symmetric, and we solve only for the azimuthal order $\nu=0$ in cylindrical coordinates. The dimensions and computational conditions are summarized in Section~\ref{sec:experiment_conditions}. The three-dimensional integral can be replaced by a weighted integral over the meridian-plane domain $\Omega_{rz}$:
\begin{equation}
 \int_\Omega q\,dV=2\pi\int_{\Omega_{rz}}q(r,z)r\,dr\,dz .\label{eq:axisym}
\end{equation}
Fig.~\ref{fig:model} shows the $R$--$Z$ cross section used in the numerical experiments and the three-dimensional coaxially fed monopole antenna obtained by rotating it. The far field is evaluated by a surface integral of the electromagnetic field near the antenna. The Huygens surface used for this integral is placed on the inner side of the PML so that fields within the absorbing layer are not included in the far-field transformation.

\section{Proposed Formulation of Antenna-Oriented Observables}
\label{sec:observables}

This section derives matrix representations of observables for evaluating antenna design metrics from the FEM coefficient vector. We introduce the underlying linear functionals, including the far-field amplitude, and use them to construct observables for the total radiated power and the radiated power within a prescribed solid angle. We also derive observables for dielectric and conductive losses and formulate the power balance with the port input power to verify the evaluations.

\subsection{Coefficient Vectors and Linear Functionals}
\label{sec:observable_basis}

Let $n$ be the number of degrees of freedom and $\bm x\in\mathbb C^n$ the FEM coefficient vector. From Faraday's law $\nabla\times\bm E=j\omega\bm\mu\bm H$, the magnetic-field basis corresponding to the electric-field basis is
\begin{equation}
 \bm E_j=\bm N_j,\qquad
 \bm H_j=\frac{1}{j\omega}\bm\mu^{-1}\nabla\times\bm N_j,
 \qquad \bm H_h=\sum_jx_j\bm H_j .
 \label{eq:magnetic_basis}
\end{equation}
By linearity, a complex linear functional $\mathcal L$ of the electromagnetic field satisfies
\begin{equation}
 \begin{aligned}
 \mathcal L(\bm E_h,\bm H_h)
 &=\sum_{j=1}^{n}\mathcal L(\bm E_j,\bm H_j)x_j
 =\bm c^{\dagger}\bm x,\\
 c_j&=\mathcal L(\bm E_j,\bm H_j)^* .
 \end{aligned}
 \label{eq:linear_functional_basis}
\end{equation}
The complex conjugate in $c_j$ ensures that $\bm c^{\dagger}\bm x$ reproduces the linear combination of the functional values on the basis fields. In the following, we obtain these coefficients explicitly from integrals of the basis fields.

\subsection{Far-Field Linear Functionals}

To construct the radiated-power observables, we express the far-field amplitude as a linear functional of the FEM coefficient vector. Let $\Gamma_H$ be the closed surface on the inner side of the PML, $\bm n_H$ its outward unit normal, $\bm J_H=\bm n_H\times\bm H$ the Huygens equivalent surface electric current, and $\bm M_H=-\bm n_H\times\bm E$ the magnetic current. Let $k$ be the wavenumber of the external medium, $\eta$ its wave impedance, $\widehat{\bm r}$ the unit vector in the observation direction, and $\bm r'$ the position on the surface. The external medium is assumed to be lossless, and the far electric field corresponding to the time-harmonic convention $e^{-j\omega t}$ is written as
\begin{equation}
 \bm E(\bm r)=\frac{e^{jkr}}{r}\bm F(\widehat{\bm r})+O(r^{-2}).
 \label{eq:farfield_convention}
\end{equation}
The coefficient $\bm F$ of this asymptotic form is given by the equivalent surface electric and magnetic currents as
\begin{equation}
\bm F(\widehat{\bm r})=\frac{jk}{4\pi}\,\widehat{\bm r}\times\left[\eta\left(\bm{\mathcal N}\times\widehat{\bm r}\right)-\bm{\mathcal L}\right], \label{eq:farfield}
\end{equation}
\begin{equation*}
 \bm{\mathcal N}=\int_{\Gamma_H}\bm J_H e^{-jk\widehat{\bm r}\cdot\bm r'}dS',\qquad
 \bm{\mathcal L}=\int_{\Gamma_H}\bm M_H e^{-jk\widehat{\bm r}\cdot\bm r'}dS' .
 \end{equation*}
By substituting the basis fields of (\ref{eq:magnetic_basis}) and projecting onto the spherical-coordinate basis $(\widehat{\bm e}_\theta,\widehat{\bm e}_\phi)$, we can precompute the coefficients using the following surface integrals. Here $\theta$ is the polar angle measured from the $+z$ axis, which is the symmetry axis, and $\phi$ is the azimuthal angle.
\begin{equation}
 \begin{aligned}
 F_\alpha(\widehat{\bm r})&=\bm t_\alpha(\widehat{\bm r})^{\dagger}\bm x,\\
 [\bm t_\alpha(\widehat{\bm r})]_j
 &=\left\{\frac{jk}{4\pi}\int_{\Gamma_H}
 \widehat{\bm e}_\alpha\cdot\left[\widehat{\bm r}\times\left(
 \eta(\bm n_H\times\bm H_j)\times\widehat{\bm r}\right.\right.\right.\\
 &\qquad\left.\left.\left.{}+\bm n_H\times\bm N_j
 \right)\right]e^{-jk\widehat{\bm r}\cdot\bm r'}\,dS'\right\}^*,
 \quad \alpha\in\{\theta,\phi\} .
 \end{aligned}
 \label{eq:farfield_linear_functional}
\end{equation}
Thus, the far-field amplitude for a specific direction and polarization is a linear functional of the coefficient vector. This expression is consistent with prior work \cite{Clader2013} in which the far-field amplitude is expressed as a linear functional and the scattering cross section is evaluated from its squared magnitude.

\subsection{Total and Solid-Angle-Restricted Radiated Power}
\label{sec:radiated_power}

The far-field amplitudes obtained in the previous subsection determine the radiation intensity, which we integrate over the observation directions to construct the power observables. The radiation intensity in the direction $\widehat{\bm r}$ is
\begin{equation}
 U(\widehat{\bm r})=\frac{|F_\theta|^2+|F_\phi|^2}{2\eta}. \label{eq:intensity}
\end{equation}
We define the evaluation matrix $T$ for the two polarizations by $T_{\alpha j}=[\bm t_\alpha]_j^*$, i.e., so that $T\bm x=(F_\theta,F_\phi)^{\mathsf T}$. Then the $ij$ component of the matrix representation of the observable is
\begin{equation}
 \begin{aligned}
 \relax [Q_U(\widehat{\bm r})]_{ij}
 &=\frac{1}{2\eta}\sum_{\alpha\in\{\theta,\phi\}}
 T_{\alpha i}(\widehat{\bm r})^*T_{\alpha j}(\widehat{\bm r}),\\
 Q_U(\widehat{\bm r})&=\frac{T(\widehat{\bm r})^{\dagger}T(\widehat{\bm r})}{2\eta},
 \qquad U=\bm x^{\dagger}Q_U\bm x .
 \end{aligned}
 \label{eq:angle_observable}
\end{equation}
Since $Q_U=T^{\dagger}T/(2\eta)$ with $\eta>0$, $Q_U$ is Hermitian positive semidefinite, consistent with the nonnegative radiation intensity. For a region $\mathcal R$ of arbitrary observation directions ($4\pi$ for the full sphere), we obtain
\begin{equation}
 \begin{aligned}
 \relax [Q_{\mathcal R}]_{ij}&=\int_{\mathcal R}[Q_U(\widehat{\bm r})]_{ij}\,d\Omega,\\
 P_{\rm rad}(\mathcal R)&=\int_{\mathcal R}U\,d\Omega
 =\bm x^{\dagger}Q_{\mathcal R}\bm x,\\
 Q_{\mathcal R}&=\int_{\mathcal R}Q_U\,d\Omega .
 \end{aligned}
 \label{eq:solid_angle}
\end{equation}
$Q_{\mathcal R}$ is also Hermitian positive semidefinite, since it is the integral of $Q_U$ over $\mathcal R$. For an antenna in free space, $\mathcal R=4\pi$ gives the total radiated power. For the monopole antenna on a PEC ground plane considered in this work, we integrate over the upper half-space $\mathcal R_+=\{0\leq\theta\leq\pi/2,\ 0\leq\phi<2\pi\}$, which is the physical radiation region. By axisymmetry, the total radiated power is
\begin{equation}
 P_{\rm rad}=P_{\rm rad}(\mathcal R_+)=2\pi\int_0^{\pi/2}U(\theta)\sin\theta\,d\theta .
 \label{eq:upper_half_power}
\end{equation}
By specifying a solid angle in (\ref{eq:solid_angle}), we can represent directional selectivity with a single observable, which the total radiated power alone cannot provide.

\subsection{Power Through the Huygens Surface and Losses}

To evaluate the total radiated power independently of the far-field integral, we define the net power passing through the Huygens surface as the surface integral of the time-averaged Poynting vector:
\begin{equation}
 P_H=\frac12\operatorname{Re}\int_{\Gamma_H}(\bm E\times\bm H^*)\cdot\bm n_HdS .
 \label{eq:huygens_power}
\end{equation}
Expanding the electromagnetic field in the basis fields of (\ref{eq:magnetic_basis}) and setting $[C_H]_{ij}=\int_{\Gamma_H}(\bm E_i\times\bm H_j^*)\cdot\bm n_HdS$, we have $P_H=\frac12\operatorname{Re}\sum_{i,j}x_ix_j^*[C_H]_{ij}$. Therefore, we obtain
\begin{equation}
 \begin{aligned}
 \relax [Q_H]_{ij}&=\frac14\left([C_H]_{ji}+[C_H]_{ij}^*\right),\\
 P_H&=\bm x^{\dagger}Q_H\bm x,\qquad Q_H=Q_H^{\dagger} .
 \end{aligned}
 \label{eq:huygens_matrix_component}
\end{equation}
The matrix $Q_H$ is Hermitian but need not be positive semidefinite for arbitrary coefficient vectors: it represents a signed net power flux through cross terms between the electric and magnetic fields. For a radiating solution with a source-free, lossless exterior, $P_H$ coincides with the nonnegative total radiated power. This property of the physical solution does not imply positive semidefiniteness of $Q_H$ over the entire coefficient space. The numerical experiments therefore assess the agreement between the power obtained from the fields on the surface and that obtained from the far-field integral.

To evaluate the components of the input power that are dissipated rather than radiated, we also express the dielectric loss and the loss at the conductor surface as quadratic forms. With the time-harmonic convention $e^{-j\omega t}$, the time-averaged dissipated powers are
\begin{equation}
 \begin{aligned}
 P_{\rm diel}&=\frac{\omega}{2}\int_{\Omega_{\rm diel}}
 \bm E_h^*\cdot\bm\varepsilon_{\rm loss}\bm E_h\,dV,\\
 P_{\rm cond}&=\frac{R_s}{2}\int_{\Gamma_s}|\bm H_{h,t}|^2\,dS\\
 &=\frac{R_s}{2|Z_s|^2}\int_{\Gamma_s}|\bm E_{h,t}|^2\,dS .
 \end{aligned}
 \label{eq:loss_fields}
\end{equation}
Here $\Omega_{\rm diel}$ is the physical dielectric region and excludes the artificial absorption of the PML. The last equality follows from $|\bm E_{h,t}|^2=|Z_s|^2|\bm H_{h,t}|^2$ under the SIBC in (\ref{eq:sibc}), for $Z_s\ne0$. Substituting the FEM basis expansion into these integrals gives
\begin{equation}
 \begin{aligned}
 P_{\rm diel}
 &=\sum_{i,j}x_i^*x_j\frac{\omega}{2}
 \int_{\Omega_{\rm diel}}\bm N_i^*\cdot\bm\varepsilon_{\rm loss}\bm N_j\,dV,\\
 P_{\rm cond}
 &=\sum_{i,j}x_i^*x_j\frac{R_s}{2|Z_s|^2}
 \int_{\Gamma_s}\bm N_{i,t}^*\cdot\bm N_{j,t}\,dS .
 \end{aligned}
 \label{eq:loss_basis_expansion}
\end{equation}
Thus, the coefficients of $x_i^*x_j$ define the loss matrices as
\begin{equation}
\begin{aligned}
 \relax [Q_{\rm diel}]_{ij}&=\frac{\omega}{2}\int_{\Omega_{\rm diel}}
 \bm N_i^*\cdot\bm\varepsilon_{\rm loss}\bm N_j\,dV,\\
 [Q_{\rm cond}]_{ij}&=\frac{R_s}{2|Z_s|^2}\int_{\Gamma_s}
 \bm N_{i,t}^*\cdot\bm N_{j,t}\,dS,\\
 P_{\rm diel}&=\bm x^{\dagger}Q_{\rm diel}\bm x,\qquad
 P_{\rm cond}=\bm x^{\dagger}Q_{\rm cond}\bm x .
\end{aligned}\label{eq:loss_integrals}
\end{equation}
For a passive dielectric with $\bm\varepsilon_{\rm loss}\succeq0$, $Q_{\rm diel}$ is Hermitian positive semidefinite. Likewise, $Q_{\rm cond}$ is Hermitian positive semidefinite for $R_s\geq0$ and $Z_s\ne0$. These properties express the nonnegativity of the respective dissipated powers. For PEC, the conductive loss is set to zero without substituting $Z_s=0$ into the SIBC loss expressions.

\subsection{Port Input Power and Power Balance}

The input power used to verify the power balance is obtained from the port reflection coefficient. Let $b_1$ be the reflected amplitude of the fundamental TEM mode and $v_1=a_1+b_1$ the total amplitude. Let $\bm h_1$ be the tangential magnetic field of the mode and $\bm s$ the orientation of the port surface. Mode projection then gives
\begin{equation}
 \begin{aligned}
 d_1&=\int_{\Gamma_p}(\bm e_1\times\bm h_1^*)\cdot\bm s\,dS,\\
 v_1&=d_1^{-1}\int_{\Gamma_p}(\bm E_t\times\bm h_1^*)\cdot\bm s\,dS
      =\bm p_1^{\dagger}\bm x,\\
 [\bm p_1]_j&=\left[d_1^{-1}
 \int_{\Gamma_p}(\bm N_{j,t}\times\bm h_1^*)\cdot\bm s\,dS\right]^*,\\
 S_{11}&=\frac{b_1}{a_1}=\frac{v_1}{a_1}-1 .
 \end{aligned}
 \label{eq:s11}
\end{equation}
The port projection $\bm p_1^{\dagger}\bm x$ is a linear functional that yields a complex modal amplitude. Its evaluation therefore differs from that of the quadratic forms for radiated and dissipated powers. Through $S_{11}$, it provides the input power used to verify the power balance of the proposed observables. The phase reference required to evaluate this complex amplitude from the solution state is discussed in Section~\ref{sec:state_evaluation}.

Let $P_{\rm inc,port}$ denote the incident power at the port. We define the ``antenna terminal input power'' $P_{\rm in,ant}$ as the power supplied to the antenna system after subtracting the reflected power. Then
\begin{equation}
 P_{\rm in,ant}=P_{\rm inc,port}(1-|S_{11}|^2)
 =P_{\rm diel}+P_{\rm cond}+P_{\rm rad,H}+R .
 \label{eq:power_balance}
\end{equation}
Here $P_{\rm rad,H}$ is the power obtained by applying (\ref{eq:huygens_power}) to the physical radiation surface, and $R$ is the power-balance residual due to numerical integration, PML truncation, and discretization. $P_{\rm inc,port}$ is determined from the normalization of the incident TEM mode.

\section{Evaluation of Observables Using a QLSS}
\label{sec:workflow}
\subsection{Preconditioning and QLSS Input}

The observables in the previous section were defined for the original FEM coefficient vector $\bm x$. Preparing the problem for a QLSS requires matrix scaling and Hermitian embedding of the non-Hermitian matrix, together with corresponding transformations of the observables. To reduce disparities in row and column scales and improve the condition number, we apply equilibration by diagonal scaling (Ruiz equilibration) \cite{Ruiz2001} to the matrix in (\ref{eq:linear_system}). The number of iterations and the change in the condition number are reported in Section~\ref{sec:experiment_conditions}. Let $D_r,D_c$ be positive real diagonal matrices, and let $A_{\rm eq}$ and $\bm b_{\rm eq}$ be the equilibrated matrix and right-hand side supplied to the QLSS. We set
\begin{equation}
\begin{aligned}
 A_{\rm eq}&=D_rAD_c,\quad
 \bm b_{\rm eq}=D_r\bm b,\quad \bm x=D_c\bm z .
\end{aligned}\label{eq:embedding}
\end{equation}
Here $A\in\mathbb C^{n\times n}$ and $\bm b,\bm x,\bm z\in\mathbb C^n$, and solving $A_{\rm eq}\bm z=\bm b_{\rm eq}$ recovers the original solution as $\bm x=D_c\bm z$. The linear functional $\bm p^{\dagger}\bm x$ is transformed as
\begin{equation}
 \bm p_{\rm eq}=D_c^{\dagger}\bm p,\qquad
 \bm p^{\dagger}\bm x=\bm p_{\rm eq}^{\dagger}\bm z . \label{eq:linear_target_transform}
\end{equation}
The quadratic form $\bm x^{\dagger}Q\bm x$ becomes $\bm z^{\dagger}Q_{\rm eq}\bm z$ with $Q_{\rm eq}=D_c^{\dagger}QD_c$. For use with an HHL-type QLSS, we construct the following Hermitian embedding of the equilibrated matrix:
\begin{equation}
 A_{\rm emb}=\begin{bmatrix}0&A_{\rm eq}\\A_{\rm eq}^{\dagger}&0\end{bmatrix},\qquad
 \bm b_{\rm emb}=[\bm b_{\rm eq};0] .
 \label{eq:hermitian_embedding}
\end{equation}
If $A$ is nonsingular, the solution of the unnormalized system $A_{\rm emb}[\bm 0;\bm z]=\bm b_{\rm emb}$ is
\begin{equation}
 \begin{bmatrix}\bm 0\\ \bm z\end{bmatrix},\qquad
 \bm z=A_{\rm eq}^{-1}\bm b_{\rm eq},
 \label{eq:embedded_solution}
\end{equation}
and the lower block of the QLSS solution state corresponds to the normalized complex solution. Therefore, the dimension after Hermitian embedding is $2n$. The corresponding observable on the system register is
\begin{equation}
 Q_{\rm emb}=\begin{bmatrix}0&0\\0&Q_{\rm eq}\end{bmatrix},
 \qquad |z_{\rm emb}\rangle=\begin{bmatrix}0\\|z\rangle\end{bmatrix},
 \label{eq:embedded_observable}
\end{equation}
where $|z\rangle=\bm z/\|\bm z\|$. Consequently, $\langle z_{\rm emb}|Q_{\rm emb}|z_{\rm emb}\rangle=\langle z|Q_{\rm eq}|z\rangle$, which connects evaluation on the embedded state to the lower-block expectation value used below.

\subsection{Observable Evaluation from the Quantum State}
\label{sec:state_evaluation}

Using HHL-type phase estimation and eigenvalue inversion, we prepare the solution state corresponding to the lower block of (\ref{eq:hermitian_embedding}). With finite phase precision, the eigenvalues are quantized. Bins for which the eigenvalue estimate obtained from the phase bin is zero, or its absolute value is below a prescribed threshold, are excluded from the inversion (eigenvalue cutoff). The approximate solution including these effects is denoted by $\bm z_{\rm HHL}$. The numerical experiments in this work classically evaluate the ideal state of the QLSS with a noiseless statevector simulator.

To recover the power on its original scale, the expectation value of the normalized state must be multiplied by the square of the solution norm. This norm is recovered from the HHL postselection success probability. In this experiment, the eigenvalues of $A_{\rm emb}$ are computed classically, and the matrix is normalized by $s_A=\max_i|\lambda_i(A_{\rm emb})|$. The amplitude of the auxiliary flag $1$ for the component with eigenvalue $\lambda$ is set to $C_{\rm inv}/\lambda$. $C_{\rm inv}>0$ is chosen so that the absolute value of this amplitude is at most 1 for the eigenvalue estimates subject to inversion. Writing $C_{\rm rot}$ for the rotation coefficient corresponding to the normalized eigenvalue $\widetilde\lambda=\lambda/s_A$, we have
\begin{equation}
 C_{\rm inv}=s_A C_{\rm rot},\qquad
 \frac{C_{\rm inv}}{\lambda}=\frac{C_{\rm rot}}{\widetilde\lambda} .
 \label{eq:inverse_scale}
\end{equation}
If the eigenvalue inversion is exact, the unnormalized amplitude vector of the success branch for the normalized input $\bm b_{\rm emb}/\|\bm b_{\rm eq}\|$ is
\begin{equation}
 C_{\rm inv}A_{\rm emb}^{-1}\frac{\bm b_{\rm emb}}{\|\bm b_{\rm eq}\|}
 =\frac{C_{\rm inv}}{\|\bm b_{\rm eq}\|}
 \begin{bmatrix}\bm 0\\\bm z\end{bmatrix} .
 \label{eq:success_amplitude}
\end{equation}
With finite phase precision, $\bm z_{\rm HHL}$ is defined as the amplitude vector of the branch in which the phase register is entirely zero, the auxiliary flag is 1, and the system register is in the solution block, divided by the same coefficient $C_{\rm inv}/\|\bm b_{\rm eq}\|$. Let $p_{\rm sol}$ denote the success probability of this joint postselection. Since the squared norm of the amplitude vector is $p_{\rm sol}$, we recover the approximate solution norm as
\begin{equation}
 \begin{aligned}
 \|\bm z_{\rm HHL}\|&=\|\bm b_{\rm eq}\|\frac{\sqrt{p_{\rm sol}}}{C_{\rm inv}},\\
 \bm x_{\rm HHL}&=D_c\bm z_{\rm HHL} .
 \end{aligned}
 \label{eq:norm_recovery}
\end{equation}
For the normalized state $|z\rangle=\bm z/\|\bm z\|$, the quadratic form is
\begin{equation}
 \bm x^{\dagger}Q\bm x=\|\bm z\|^2\langle z|Q_{\rm eq}|z\rangle .
 \label{eq:physical_expectation}
\end{equation}
In the QLSS evaluation, $\bm z$ is replaced by $\bm z_{\rm HHL}$. The linear functional can also be obtained as $\bm p^{\dagger}\bm x=\|\bm z\|\,\bm p_{\rm eq}^{\dagger}|z\rangle$, but evaluating the complex amplitude requires a phase reference relative to the known incident amplitude. In this experiment, these expectation values and complex amplitudes are evaluated on the simulator.

\section{Numerical Experiments}
\label{sec:numerical_results}
\subsection{Experimental Conditions}
\label{sec:experiment_conditions}

We compare the classical FEM direct solver and the HHL algorithm in two sets of numerical experiments: (i) a single-frequency comparison of radiated power using a lossless PEC model and (ii) a 21-point frequency sweep of the power balance.

We use normalized vacuum units: the wave impedance of the external air medium is $\eta_0=1$, the permeability is $\mu=1$, the permittivity is $\varepsilon_{\rm air}=1$, and the real relative permittivity of the coaxial dielectric is $\varepsilon_{r,\rm coax}=2.1$. The normalized frequency $f$ satisfies $\omega=2\pi f$, and the wavenumber of the external air is $k=\omega$. For a reference length $L_0$ and the speed of light in vacuum $c_0$, the corresponding physical frequency is $f_{\rm phys}=fc_0/L_0$. In this work, the physical unit of $L_0$ is not specified, and dimensions, power, and surface impedance are reported in normalized units. The real relative permittivity of the coaxial dielectric is used to normalize the incident TEM mode.

The model dimensions are as follows: the inner radius of the coaxial conductor is $0.04$, the outer radius is $0.14$, the coaxial length is $0.3$, the monopole length is $0.55$, and the air region has a radius of $1.2$ and a height of $1.45$, with a PML of thickness $0.5$ placed outside. The PML has a quadratic profile with a dimensionless strength parameter of $7$.

The radiated-power comparison is performed at $f=0.4$. This standard case has 519 complex degrees of freedom, a Hermitian embedding of dimension $1,038$, and a dimension of $2,048$ after zero padding (11 qubits for the system register). The far field is evaluated for two polarizations at 91 points in $1^\circ$ increments over $\theta=0$--$90.0^\circ$, and the total radiated power is obtained from (\ref{eq:upper_half_power}). To verify the evaluation over a prescribed solid angle, we use an annular band $42.0^\circ\leq\theta\leq48.0^\circ$, $0\leq\phi<2\pi$, with 7 quadrature points. To obtain a far field that satisfies the PEC ground-plane boundary condition, the fields on the upper air--PML surface are transformed after adding their image with respect to the ground plane. Adding the image of the 105 surface points gives 210 evaluation points. In contrast, $P_{\rm rad,H}$ is the power passing through the physical upper air--PML surface, without the image contribution. This surface is also used to evaluate the power balance with a finite surface impedance, without the PEC image.

The power balance is evaluated over a 21-point frequency sweep at $f=0.30$--$0.50$. In addition to the lossless condition, a high-loss condition with $\epsilon_{\rm loss,coax}=0.2$ and $R_s=X_s=0.001$ is used to illustrate power dissipation. Here $\epsilon_{\rm loss,coax}$ is the isotropic scalar coefficient of $\bm\varepsilon_{\rm loss}$ assigned to the coaxial dielectric; the loss coefficient is zero in the air region.

For both the single-frequency radiation verification and the power-balance sweep, the phase register had $q_{\rm phase}=14$ qubits. The number of Ruiz iterations was 8 in all cases. Since the eigenvalue $\lambda$ is mapped to the phase $0.49\lambda/s_A$, the eigenvalue resolution is
\begin{equation}
 \Delta\lambda=\frac{s_A}{0.49 \cdot 2^{q_{\rm phase}}} .
 \label{eq:phase_resolution}
\end{equation}
The cutoff is set to $\Delta\lambda$, and bins for which the estimated eigenvalue $\widehat\lambda$ obtained from the phase bin for the embedded and zero-padded matrix is zero or satisfies $|\widehat\lambda|<\Delta\lambda$ are excluded from the inversion. The rotation coefficient is $C_{\rm inv}=0.9\Delta\lambda$, i.e., $C_{\rm rot}=0.9/(0.49\cdot2^{q_{\rm phase}})$, which keeps the rotation amplitude of the retained bins at an absolute value of at most $0.9$. In the sweep, the cutoff and $C_{\rm inv}$ are recomputed from $s_A$ for each frequency and material condition, and $p_{\rm sol}$ is also evaluated at each point. No physical components of the right-hand side or the matrix representations of the observables are placed in the zero-padded part.

\begin{figure}[t]
  \centering
  \includegraphics[width=.86\columnwidth]{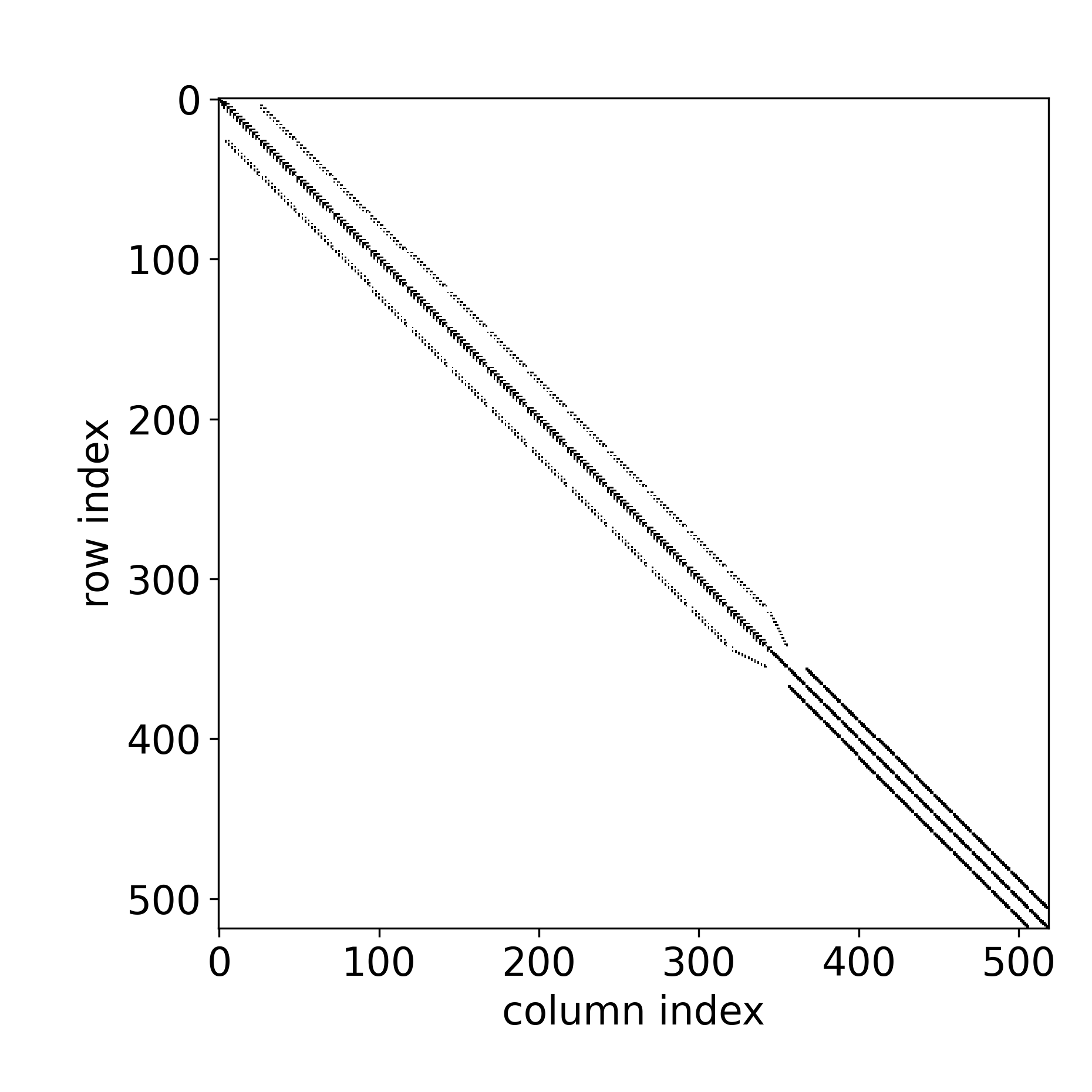}\\[-1.5mm]
  \includegraphics[width=.86\columnwidth]{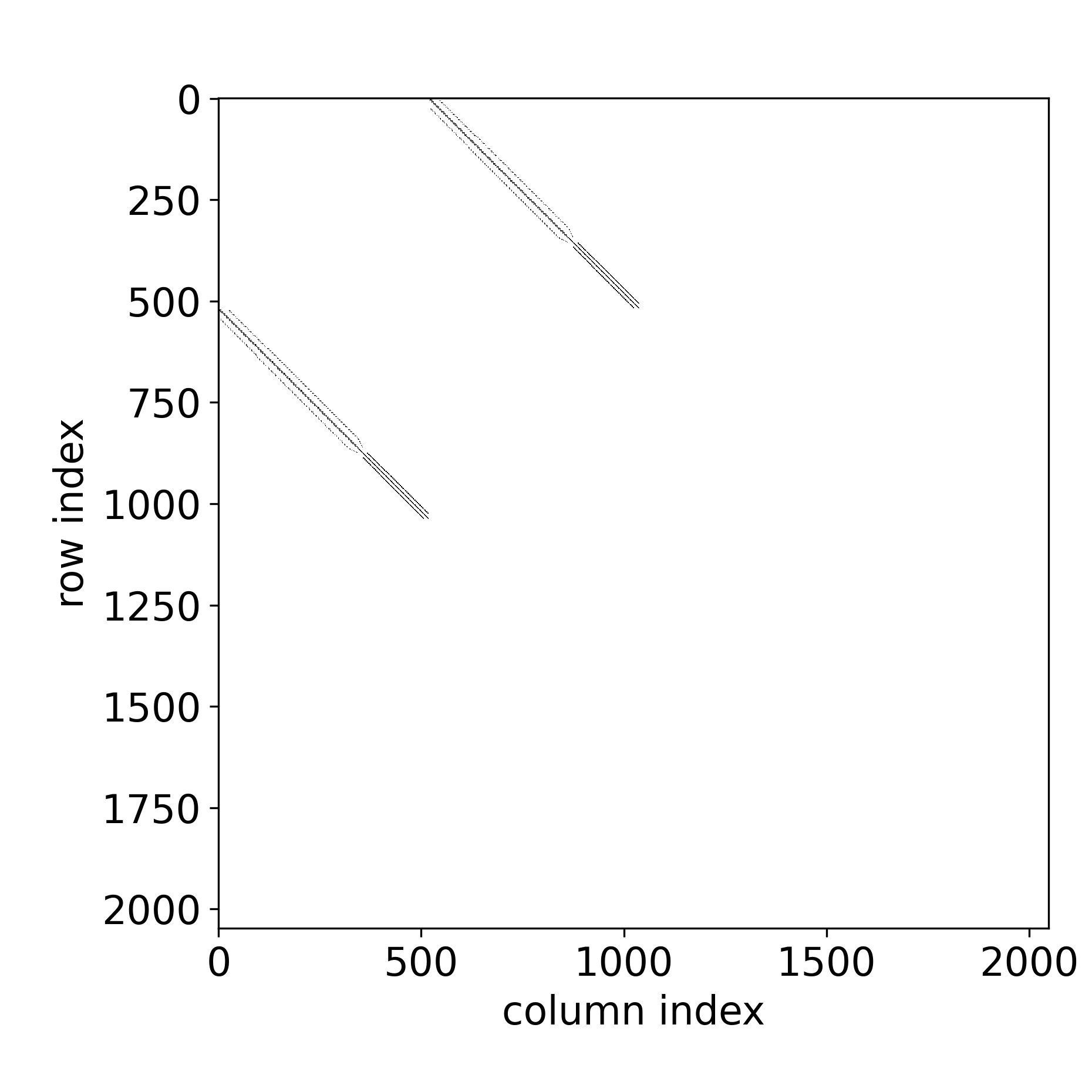}
  \caption{Matrix sparsity patterns of the experimental system (black indicates $|A_{ij}|>0$). The upper panel shows the complex FEM matrix ($519\times519$), and the lower panel shows the quantum input matrix zero-padded to $2,048\times2,048$ after Hermitian embedding.}
  \label{fig:sparsity}
\end{figure}

Fig.~\ref{fig:sparsity} shows the matrix sparsity pattern for the standard case. The complex matrix before Hermitian embedding has 2--9 nonzero entries per row (mean 6.94), and the sparsity arising from the local basis functions is preserved after embedding. For the excitation in this experiment, only one component of the right-hand side was nonzero. For the complex $519\times519$ matrix of the standard case, the 2-norm condition number, given by the ratio of the largest to the smallest singular value, decreased from $\kappa_2(A)=6.16\times10^5$ to $\kappa_2(A_{\rm eq})=9.34\times10^3$ after eight Ruiz equilibration iterations. In the single-frequency radiation verification at $f=0.4$, the joint postselection success probability used for norm recovery in (\ref{eq:norm_recovery}) was $p_{\rm sol}=1.29\times10^{-5}$.

\begin{figure*}[t]
  \centering
  \includegraphics[width=.93\textwidth]{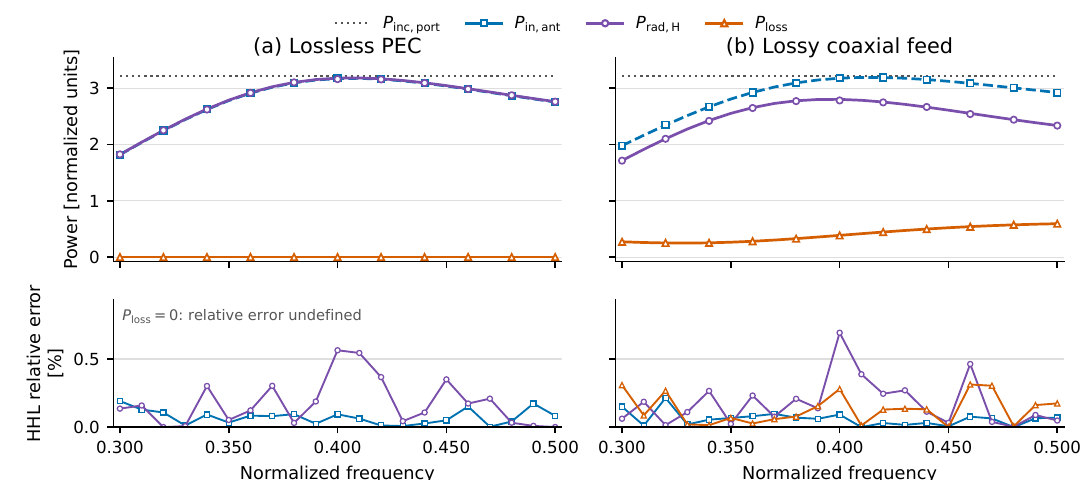}
  \caption{Frequency sweeps for the coaxially fed model under (a) lossless PEC and (b) high-loss conditions. The upper panels show power, and the lower panels show the relative error of the HHL algorithm with respect to the direct solver for each power quantity. Blue indicates the antenna terminal input power, purple the radiated power through the Huygens surface, and orange the sum of the dielectric and conductive losses. In the upper panels, curves show the direct-solver results, open markers show the HHL results (only every other point is displayed), and the gray dotted line shows the incident power common to both methods. Relative errors are shown for all 21 points; they are undefined for the dissipated power in the lossless case because that power is zero. The high-loss condition is $\epsilon_{\rm loss,coax}=0.2$, $R_s=X_s=0.001$.}
  \label{fig:power_sweep}
\end{figure*}

\subsection{Far-Field and Solid-Angle-Restricted Power}

At a normalized frequency of $0.4$, we compared the classical FEM and QLSS evaluations. The power quantity under evaluation is denoted by $V$, and the relative error is defined as $\delta_{\rm rel}=|V_{\rm HHL}-V_{\rm direct}|/|V_{\rm direct}|$. Errors are computed from unrounded values, and the power values in Table~\ref{tab:metrics} are rounded to three significant digits. As shown in the table, the relative errors between the direct solver and the HHL algorithm were below $10^{-2}$ for the total radiated power, the annular-band power, and the Huygens-surface power. The difference between the power evaluated using the squared magnitude of the far-field linear functional and that obtained by directly evaluating the corresponding quadratic form was at most $2.00\times10^{-15}$ for the classical solution and $4.33\times10^{-15}$ for the QLSS solution. These results confirm that both the total radiated power and the power within a prescribed solid angle can be evaluated as observables using the same solution.

\begin{table}[t]
  \caption{Comparison of radiated power at a normalized frequency of $0.4$. Power is reported in the normalized units defined in Section~\ref{sec:experiment_conditions}.
}
  \vspace{1mm}
  \label{tab:metrics}
  \centering\scriptsize
  \resizebox{\columnwidth}{!}{%
  \begin{tabular}{lccc}
    \hline
    Power & Direct solver & HHL algorithm & Relative error\\
    \hline
    Total radiated power & $3.39$ & $3.37$ & $7.69\times10^{-3}$\\
    Annular-band power & $0.164$ & $0.163$ & $6.62\times10^{-3}$\\
    Huygens-surface power & $3.18$ & $3.17$ & $4.87\times10^{-3}$\\
    \hline
  \end{tabular}%
  }
\end{table}

Even for the classical solution, however, a difference of about $6.64\%$ remains between the radiated power obtained from the far-field integral and the power passing through the Huygens surface, with the latter as the reference. Agreement between the QLSS and the classical solution alone therefore cannot guarantee the accuracy of the electromagnetic analysis itself.

\subsection{Power Balance Including Losses}

In the model with losses, the terminal input power, which excludes reflected power, is divided between radiation and material losses. We verify the power balance in (\ref{eq:power_balance}) and compare the dissipated powers evaluated by the QLSS with those obtained from the classical solution.

Fig.~\ref{fig:power_sweep} shows the results of the frequency sweep. In the lossless case, the curves for terminal input power and radiated power nearly coincide, whereas under the high-loss condition, part of the input power is dissipated as $P_{\rm loss}=P_{\rm diel}+P_{\rm cond}$.

Across all 21 points, the relative errors of the HHL algorithm with respect to the direct solver were at most $0.564\%$ (lossless) and $0.693\%$ (high loss) for the radiated power, and at most $0.313\%$ for the total loss under the high-loss condition. The maximum relative errors in the individually evaluated dielectric and conductive losses were $1.23\%$ and $0.580\%$, respectively. Each loss observable was consistent with the classical solution of the same discrete equation.

\section{Discussion}
\label{sec:discussion}

For this model, we confirmed that the power values evaluated by the QLSS are close to those obtained from the classical solution of the same discrete equation. Questions remain concerning the accuracy of the electromagnetic analysis itself and the practical performance of the quantum computation.

The difference between the power obtained from the far-field integral and that passing through the Huygens surface, reported in Section~\ref{sec:numerical_results}, requires a convergence study. Such a study must separate the contributions of the mesh, numerical integration, and discretization of the integration surface to assess the accuracy of the electromagnetic analysis.

The quantum computation was evaluated using an idealized classical simulation that includes finite phase precision and the eigenvalue cutoff. We did not evaluate the synthesis precision of rotation gates or the quantum error correction resources required for implementation on fault-tolerant quantum computers (FTQCs). The expectation values and postselection success probabilities were obtained from the simulator, so estimation errors due to a finite number of measurements were also outside the scope of the evaluation.

Formulating quantities as observables and having a sparse matrix and right-hand side do not, by themselves, guarantee efficient expectation-value estimation. Given the cost of expectation-value estimation \cite{Alase2022}, the total computational complexity from state preparation to measurement must be evaluated. This evaluation must include the cost of obtaining a spectral upper bound to replace $s_A$, which was computed classically in this experiment. The overall complexity of a QLSS depends not only on matrix size but also on the condition number, matrix access (block-encoding), state preparation, and the cost of estimating the observables. Detailed resource estimates and end-to-end assessments of quantum radar-cross-section calculations also indicate large circuit costs \cite{Scherer2017,Parker2026}.

Efficient matrix access and preconditioning also require attention: relevant approaches include high-precision elliptic PDE solvers \cite{ChildsLiuOstrander2021}, preconditioned quantum FEM \cite{DeimlPeterseim2026}, quantum assembly \cite{Alkadri2025}, fast inversion \cite{TongEtAl2021}, and domain decomposition for Poisson FEM \cite{Fressart2026}. Their applicability to the complex indefinite Maxwell system considered here requires separate analysis.

Future resource estimates should consider solvers with optimal condition-number and precision dependence \cite{CostaEtAl2022,Dalzell2024}, including recent optimal query bounds in the sparse-matrix oracle model \cite{BravoPrieto2026OptimalQLSA}. Instance-dependent solvers that can improve on worst-case condition-number estimates provide another option \cite{Dalzell2026BeyondConditionNumber}. Comparisons should include matrix access and power-observable estimation, rather than solution-state preparation alone.

\section{Conclusion}
\label{sec:conclusion}

We formulated observables for evaluating antenna design metrics from the coefficient vector of finite-element electromagnetic analysis. We expressed the far-field amplitude as a linear functional and used its squared magnitude to construct Hermitian matrix representations of the total radiated power and the radiated power within an arbitrary prescribed solid angle. We also derived the dielectric and conductive losses as quadratic forms of the electric field. These quantities can be compared with the terminal input power obtained from port projection using the same coefficient vector.

We presented a hybrid quantum--classical workflow that combines these observables with Ruiz equilibration, Hermitian embedding of the complex matrix, and solution-state preparation using an HHL-type QLSS. Numerical experiments with an ideal QLSS for an axisymmetric coaxially fed monopole confirmed agreement with the classical FEM solution of the same discrete equation for the total radiated power, radiated power within a prescribed solid angle, power passing through the Huygens surface, and dissipated power. These results demonstrated the consistency of the formulation for evaluating multiple global antenna design quantities from the same QLSS solution state without reconstructing all components of the solution vector.

Future work will assess the convergence of the electromagnetic analysis and quantify the total resources required for implementation and measurement on FTQCs. The formulation is expected to support the application of QLSSs to antenna design for high-frequency wireless devices, including frequency sweeps and design-parameter exploration.

\ifdefined\DIFadd\else
  \IEEEtriggeratref{24}
\fi
\bibliographystyle{IEEEtran}
\bibliography{references}

\end{document}